\documentclass{article}

\usepackage{microtype}
\usepackage{graphicx}
\usepackage{subfigure}
\usepackage{amsmath}
\usepackage{booktabs} % for professional tables
\usepackage{multicol}
\usepackage{makecell}

\usepackage{hyperref}

\usepackage[accepted]{icml2023}
\begin{document}

\twocolumn[
\icmltitle{A Method for Auto-Differentiation of the Voronoi Tessellation}




\begin{icmlauthorlist}
\icmlauthor{Sergei Shumilin}{to}
\icmlauthor{Alexander Ryabov}{to}
\icmlauthor{Serguei Barannikov}{to,fr}
%\icmlauthor{Nikolay Yavich}{to}
\icmlauthor{Evgeny Burnaev}{to}
\icmlauthor{Vladimir Vanovskii}{to}
\end{icmlauthorlist}

\icmlaffiliation{to}{Applied AI Center, Skoltech, Moscow, Russia}
\icmlaffiliation{fr}{CNRS, Université Paris Cité, France}

\icmlcorrespondingauthor{Sergei Shumilin}{ishumili@gmail.com}
\icmlcorrespondingauthor{Alexander Ryabov}{A.Ryabov@skoltech.ru}
\icmlcorrespondingauthor{Evgeny Burnaev}{burnaevevgeny@gmail.com}
\icmlcorrespondingauthor{Vladimir Vanovskii}{vladimir.vanovsky@gmail.com}

\icmlkeywords{Machine Learning}

\vskip 0.3in
]



\printAffiliationsAndNotice{}  % leave blank if no need to mention equal contribution

\begin{abstract}
We present the method for autodifferentiation of the 2D Voronoi tessellation. The method allows one to construct the Voronoi tessellation and pass the gradients, making any numerical construction involving it end-to-end differentiable. We provide the implementation details and test it on several problems, such as optimal hospital placement. To the best of our knowledge this is the first autodifferentiable realization of the Voronoi tessellation providing full set of Voronoi geometrical parameters in a differentiable way. 
\end{abstract}

\section{Introduction}
\label{submission}

%INCLUDE: https://en.wikipedia.org/wiki/Lloyd%27s_algorithm

Voronoi tessellation is ubiquitous in a broad range of science and engineering. The tessellation forms the fundamental concept in computational geometry, partitioning a given space into regions based on proximity to a set of predefined points. It offers a unique lens to examine and model diverse phenomena, from the microscopic structure of cellular tissues to the macroscopic layout of urban landscapes.

The versatility of Voronoi diagrams extends to fields such as physics \cite{kobayashi2002crystal, de2004formation, debnath2016edge}, climate modeling \cite{osti_1090872, liu2023intelligent},  biology \cite{math9212726, marcelpoil1992methods, edla2012clustering}, material science \cite{sun2018microscale, acton2018voronoi}, and geographic information systems (GIS) \cite{nowak2015application}, providing crucial insights in each domain. For instance, in biology, they help model cellular growth patterns, while in GIS, they aid in efficient land use planning. Despite their widespread utility, the application of Voronoi tessellation in optimization, especially in solving inverse problems such as those encountered in fluid dynamics, presents unique challenges. These problems often require the integration of observational data, to optimize the placement of reference points for numerical simulations. Such problems can be solved efficiently using optimization algorithms and machine learning. There are a large number of optimization algorithms, some of the most efficient of which are differentiable optimization methods. However, obtaining Voronoi tessellation characteristics, such as Voronoi region areas or adjacent edge lengths, involves the use of Voronoi tessellation algorithms, which are non-differentiable. 

In this paper, we introduce a novel method that enables automatic differentiation of the process of construction of the Voronoi tessellation. This approach promises to enhance the solution of inverse problems across various disciplines by bridging the gap between the differentiable optimization methods and the non-differentiable nature of the Voronoi tessellation algorithm.

Hence, our main contributions are as follows:

\begin{itemize}
    \item We developed a method for auto-differentiation of the Voronoi tessellation that makes possible to pass gradients from arbitrary loss function and optimize the geometry of the site points;
    \item We extend out method to the case of bounded Voronoi tessellation;
    \item We demonstrate the applicability of our method on different problems;
    %\item We published the open-source PyTorch-based version of our code as a Python package making it easy-to-use in the broad range of applications.
\end{itemize}

\subsection{Related Work}

There are several works that are considered related. 
The work \cite{diff_delaunay} presents a method to get a differentiable version of the Delaunay triangulation.
In \cite{chen2022semi} the authors combined Voronoi tessellation with normalizing flows to construct a new type of invertible transformation. This transformation facilitates applications such as Voronoi dequantization and disjoint mixture models, where discrete values are mapped into a continuous space. However, the method described therein does not make broad range of the Voronoi tessellation parameters differentiable w.r.t. input of Voronoi tessellation (point clouds).  Instead, the authors focused on integrating Voronoi tessellation into a framework for modeling the transition between discrete and continuous spaces. This integration uses Voronoi tessellation to partition space and create mappings that are amenable to gradient-based optimization. However, the process of tessellation itself, including the derivation of specific parameters like cell volumes, is not set up for direct differentiation in relation to the input data. 

\cite{feng2023cellular} explores an innovative approach to differentiable Voronoi tessellation, focusing on a "soft" Voronoi tessellation (in other words, approximation).  In contrast, our research takes a fundamentally different approach by focusing on the differentiability of "hard" Voronoi tessellation. Our method retains the explicit, well-defined boundaries of Voronoi regions, enabling precise control over their geometric properties through gradient-based optimization. This approach is especially advantageous in scenarios where exact spatial configurations are crucial, such as in complex engineering designs or high-precision modeling tasks.

\section{Background}

\begin{table}
    \centering
    \begin{tabular}{|l|l|} \hline 
        Voronoi tessellation & $VT$ \\ \hline 
        Delaunay Triangulation & $DT$ \\ \hline 
        Site points & $S$ \\ \hline 
        Voronoi diagram & $V(S)$ \\ \hline 
        Voronoi region & $VR(x, S)$ \\ \hline 
        Delaunay triangulation of $S$ & $DT(S)$ \\ \hline 
        N & $Card(S)$ \\ \hline 
        border edges of $DT(S)$ & $\partial DT(S)$ \\ \hline 
        loss function & $L$ \\ \hline
        circumcenter of a triangle & $v$  \\ \hline 
        site point & $x$ \\ \hline 
        Delaunay edge & $e$ \\ \hline 
        \makecell[l]{vector corresp. to \\ border Delaunay edge} & $\vec{d_i}$ \\ \hline
        \makecell[l]{$V(S)$ with a set \\ of Voronoi regions} & $VR^+(S)$ \\ \hline
        \makecell[l]{circumcenters of triangles \\ adjacent to a site point x} & $C(x)$ \\ \hline
        Delaunay triangles & $T$ \\ \hline
        area of a Voronoi region & $A$ \\ \hline
        set of Voronoi edges & $VE(S)$ \\ \hline
        Voronoi points & $VP(x)$ \\ \hline
        number of optimization steps & $m$ \\ \hline
    \end{tabular}
    \caption{Main definitions}
    \label{tab:my_label}
\end{table}

\subsection{Voronoi tessellation}

The input of our method is the set of points $S$ with more than three points.
We formally define Voronoi region for a given point $p$ as $VR(p, S)$ or $VR(p)$ if it's clear what $S$ is mentioned.
$VR(p)$ consists of all points in $\mathbf{R}^2$ for which $p$ is the nearest neighbour site.

\textbf{Definition 1. } \cite{Aurenhammer} \textit{The common boundary part of two Voronoi regions is called a Voronoi edge if it contains more than one point. The Voronoi diagram of S, for short V (S), is defined as the union of all Voronoi edges.}

The Voronoi points are positioned on the common boundary of three or more Voronoi regions and serve as the endpoints of Voronoi edges. We denote a set of Voronoi points for given site point x as $VP(x)$.

Obtaining the convex hull of S is straightforward by using its Delaunay triangulation.
Easily derived from the finite set S, the convex hull is the smallest possible convex set that encompasses it.
The convex hull of S is a convex polygon that is formed by h $\leq$ n extreme points of S. It possesses very useful properties related to the infinite Voronoi region which we decribe further.

\subsection{Delaunay triangulation}

Delaunay triangulation is a pivotal element of our method, as it plays a crucial role in the computation of Voronoi diagrams.

We define Delaunay triangulation of a set of site points $S$ as $DT(S)$.

\textbf{Definition 2.}\textit{ $DT(S)$ consists of polygonal faces over $S$ whose circumcircles do not contain any points from $S$.}

The main property that we leverage in the outside method is the duality between $V(S)$ and $DT(S)$.

\subsection{Auto-differentiation}

Automatic differentiation (AD) is a technique for obtaining the gradients of a program's output with respect to input  \cite{Naumann}.
One of the most important algorithms based on AD is backpropagation.
Nowadays, it's primary application is in training deep learning models.
Backpropagation is proved to be powerful enough to allow training even Large Language Models with billions of parameters.

However, AD is used in much broader sense in numerous applications in physics \cite{dold2023differentiable, liang2020differentiable}, chemistry \cite{kasim2022dqc, arrazola2021differentiable, wang2022end} and biology \cite{alquraishi2021differentiable}.
A program or a neural network can be represented as a computational graph in which nodes represent functions, and edges represent input/output relationships.
Each node of the computational graph corresponds to a simple operation and, in fact, represents a function.
Simple functions then combined to form a much complex models.
The computations graph contains nodes of three types: leaf nodes which are constants or input variables, terminal nodes which represent output, and regular nodes which are none of the others. 

There is a number of Python based frameworks for AD. For our experiments we use PyTorch \cite{paszke2017automatic}.

\begin{figure*}[hbt!]
\begin{center}
\includegraphics[width=\textwidth]{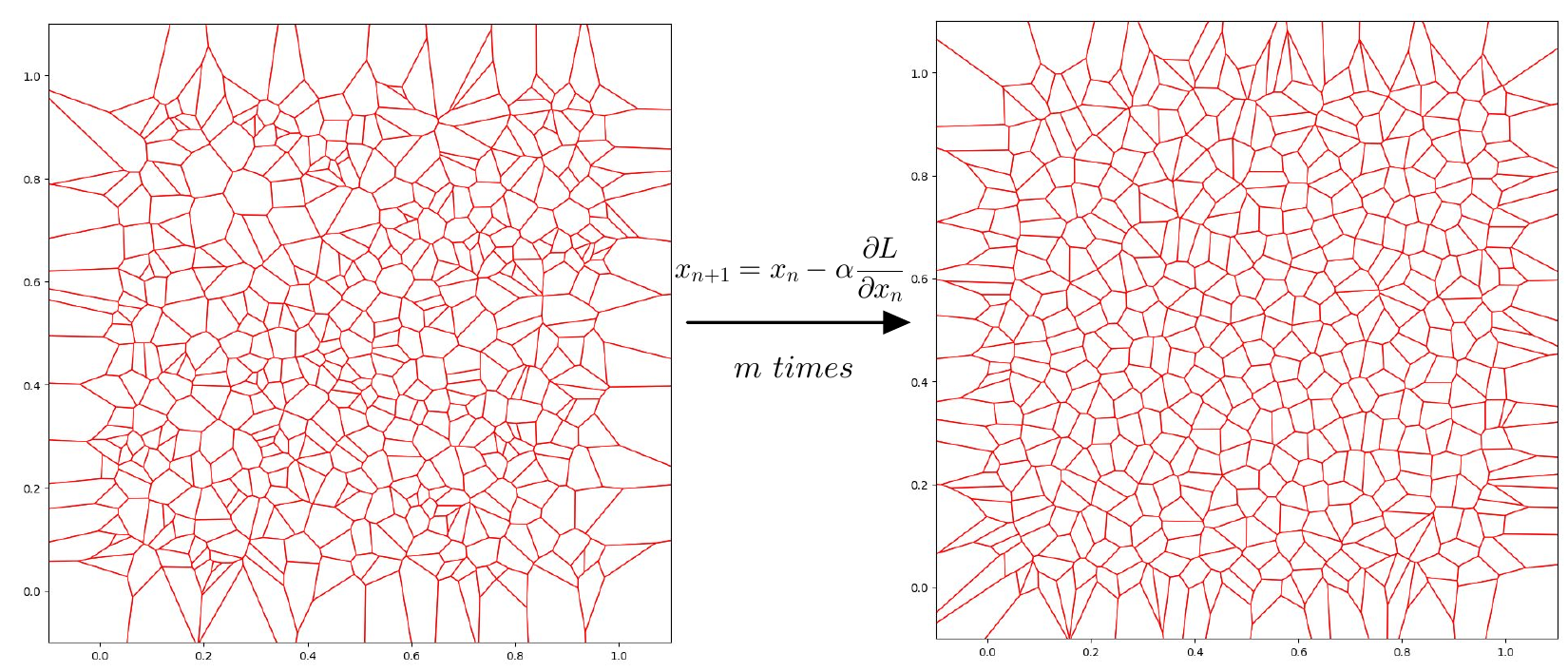}
\caption{On the left is the initial Voronoi tessellation of 500 random points. On the right the final Voronoi tessellation after optimizing the position of the site points. We iteratively optimize the placement of the site points to align the areas of the Voronoi regions. As a result we obtain the Voronoi regions with uniform areas. The Voronoi regions are shown red. Site points are not shown. There is no border so the site points may move any place. We make the infinite region finite and their corresponding site points are also optimized. We define the retriangulation step $r$.}
\label{unbounded_optimization}
\end{center}

\end{figure*}

\section{Auto-differentiable Voronoi tessellation}

For simplicity, we assume that $S$ are in general position.
First, we obtain the $DT(S)$.
We do not include the calculation of $DT(S)$ in the computational graph of AD.
We construct the computational graph dynamically based on the adjacency information obtained after $DT(S)$.
Therefore, to compute Delaunay triangulation we may choose any available software.
Because of the duality between $DT(S)$ and $V(S)$ we may calculate the geometry of $V(S)$ based on the pre-computed $DT(S)$.

\begin{figure}[hbt!]
\begin{center}
\includegraphics[width=\columnwidth]{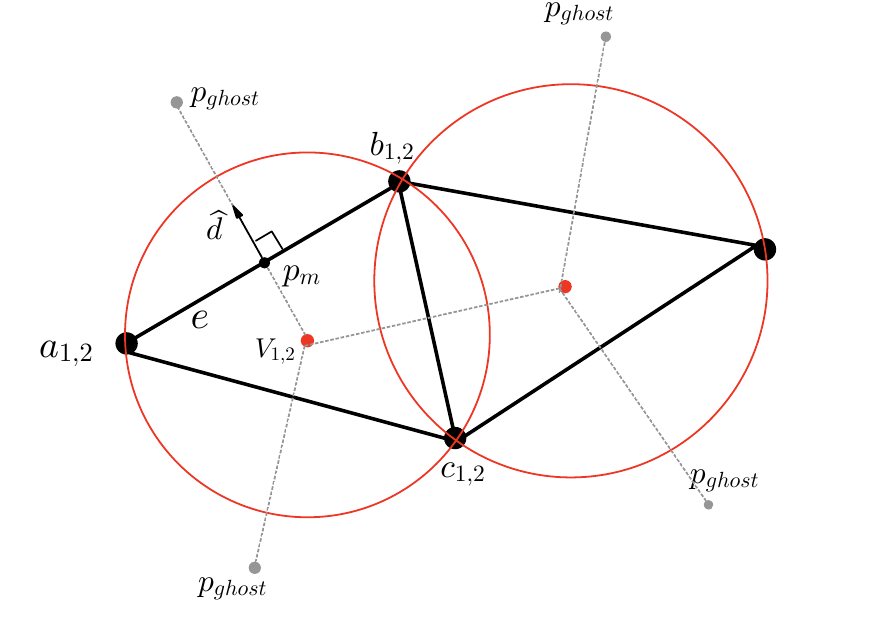}
\caption{Intersections of the Voronoi edges are the centers of circumcircles of Delaunay triangles. These are red points in the center of the circumcircles. Voronoi edges are shown in grey dashed lines. The Delaunay triangles are shown in bold black lines. For every border edge there is a ghost point that we place in order to process infinite edges and areas. For every border edge $e_i$ there is a vector $R$ that is perpendicular to $e_i$ and starts from the middle point of $e_i$. The corresponding ghost point is placed on this vector.}
\label{scheme}
\end{center}

\end{figure}

\textbf{Voronoi points.}
Let $x_i$ be a site point. Let $DT(x_i, S)$ be a set of all Delaunay triangles adjacent to $x_i$.
The Voronoi points of $VR(x_i, S)$ are circumcenters of $DT(x_i, S)$ (see Fig.~\ref{scheme}). For infinite Voronoi region we introduce ghost points which lie on infinite Voronoi edges.

Let $a_{1, 2}, b_{1, 2}, c_{1, 2}$ be the three site points that constitute a Delaunay triangle $t_{a,b,c}$.
Subscripts in $a_{1, 2}$ stand for $x$ and $y$ coordinates, respectively. 
The analytical expression to calculate the circumcenter of $t_{a,b,c}$:
\begin{multline}
v_1 = \frac{\left(a_1^2-c_1^2+a_2^2-c_2^2\right)\left(b_2-c_2\right)}{D} - \\ 
    -  \frac{\left(b_1^2-c_1^2+b_2^2-c_2^2\right)\left(a_2-c_2\right)}{D} \label{circumcenters_x} \\
\end{multline}
\begin{multline}
v_2 = \frac{\left(b_1^2-c_1^2+b_2^2-c_2^2\right)\left(a_1-c_1\right)}{D} - \\ 
    -  \frac{\left(a_1^2-c_1^2+a_2^2-c_2^2\right)\left(b_1-c_1\right)}{D} \label{circumcenters_y} \\
\end{multline}
where $D$ which is four times the area of $t_{a,b,c}$ , is given by
\begin{align}
D=2\left[\left(a_1-c_1\right)\left(b_2-c_2\right)-\left(b_1-c_1\right)\left(a_2-c_2\right)\right]
\end{align}
It's clear that we can calculate $\displaystyle  \frac{\partial v_1}{\partial a}$, $\displaystyle  \frac{\partial v_2}{\partial a}$ and the same for $b$ and $c$.

\textbf{Voronoi edges.}
Some of the Voronoi edges are of infinite length.
They correspond to infinite Voronoi regions.
In order to process infinite Voronoi edges computationally we detect them and make them finite.
We utilize the property of the Delaunay-Voronoi duality:

\textbf{Lemma 1} \cite{Aurenhammer} \textit{ A point $p$ of $S$ lies on the boundary of the convex hull of $S$ iff its Voronoi region $VR(p, S)$ is unbounded.}

Therefore we need to process $\forall e \in \partial DT(S)$ and handle infinite Voronoi edges.
Every border site point is adjacent to two border edges of $\partial DT(S)$. Note that $\partial DT(S)$ consists of border edges whereas $ConvexHull(S)$ consists of border points.

Let ${p, q} \in ConvexHull(S)$ be two border site points connected by a border edge $e_i$. 
Let $\vec{d_i} = VR(p, S) \cup VR(q, S)$ be a ray (see Fig.~\ref{scheme}).
We calculate the middle point $p_m = (p + q) / 2$. $d_i$ starts in $p_m$ and is perpendicular to $e_i$.
There are two such rays.
We choose the one that points outside of the $VT$.
We place a ghost point $p_{ghost}$ on $d_i$.
To do so we multiply normalized $d_i$ by the predefined constant $\omega$ which is a hyperparameter.
The larger $\omega$ the better approximation we obtain. Formally:
\begin{equation}
    p_{m} = 0.5 \cdot (p + q),
\end{equation}
\begin{equation}
    \vec{d_i} = [p - q]^{\perp},
\end{equation}
\begin{equation}
    \hat{d_i} = \frac{ \vec{d_i} } { | \vec{d_i} | },
\end{equation}
\begin{equation}
    p_{ghost} = p_m + \omega \cdot \hat{d_i}.
\end{equation}
Let $L(V(R), \phi)$ be an objective function where $\phi$ is set of additional parameters.
$\phi$ may represent e.g. physical quantities that may be optimized and represents a leaf node of the computational graph.
\begin{figure}[hbt!]
\begin{center}
\includegraphics[width=\columnwidth]{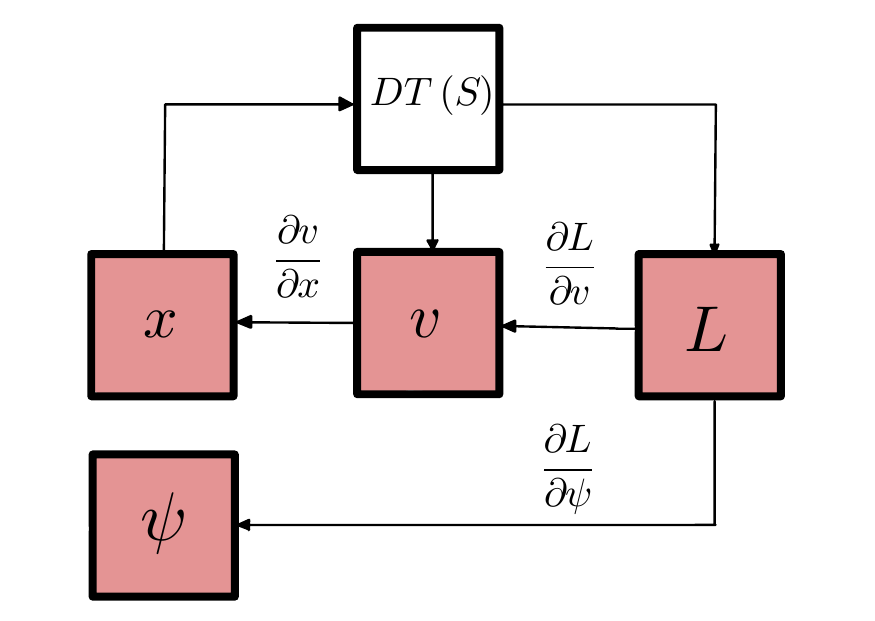}
\caption{Principled scheme of the backward pass. Red blocks are differentiable. $DT(S)$ only provides adjacency information for derivation. We show $\psi$ as a differentiable parameters, however, it can be non-differentiable. The AD will make $\psi$ a leaf-node in the computational graph as well as $x$. }
\label{scheme of differentiation}
\end{center}
\end{figure}
The chain of partial derivatives that we calculate using the backpropagation algorithm (see Fig.~\ref{scheme of differentiation}):
\begin{equation}
    \frac{\partial L}{\partial x} = \frac{\partial L}{\partial v}
    \cdot 
    \frac{\partial v}{\partial x}.
\end{equation}
The main advantage of the technique is flexibility because we may choose any $L$ function that supports autograd.
On the contrary, if we use explicit differentiation we need to code explicit derivative for every $L$.

Programming realization of our method outputs data structures containing geometric description of the $VR^+(S)$. So $L$ must be a function of $VR^+(S)$ geometry. For example it may be a function of the areas of the Voronoi regions.

\textbf{Retriangulation.} As we optimize the position of the site points $S$, $DT(S)$ evolves. 
Despite the complexity of the Delaunay triangulation is $O(N~\log N)$, doing triangulation on every step of optimization may be time consuming. 
To tackle this we set the frequency of triangulation $r > 1$. 
It controls the number of steps of optimization when particular $DT(S)_i$ is kept. 
After we have done $r$ steps we do triangulation again. 
As the positions of site points change smoothly because of the small step of gradient descent algorithm, the circumcenters also change smoothly because both \ref{circumcenters_x} and \ref{circumcenters_y} are smooth functions. 
The supporting experiment data presented in the Application section.

When each r steps of optimization passes, we use updated convex hull for detection of infinite Voronoi edges using \textbf{Lemma 1}, as described earlier. 
Let $PointCircumcenters$ be a map data structure that maps a site point $x$ to $C[x]$. 
For the pseudo code we except the dictionary-like notations for $PointCircumcenters$. Also we use array-like notation $[~]$ for $S$, $C$ and $T$.
The overall algorithm of differentiable Voronoi tessellation is described in \ref{alg:differentiable_voronoi}. 

\begin{algorithm}[hbt!]
   \caption{Differentiable Voronoi tessellation}
   \label{alg:differentiable_voronoi}
\begin{algorithmic}
   \STATE {\bfseries Input:} site points $S$, number of optimization steps $m$, frequency of retriangulation $r$
   \STATE $T$, $ConvexHull(S)$ = $DT(S)$
   \REPEAT
   \STATE Map $PointCircumcenters$

   \FOR{$i=1$ {\bfseries to} $|T|$}
   \STATE Find circumcenter of $T[i]$
   \ENDFOR

   \FOR{$i=1$ {\bfseries to} $|S|$}
   \IF{$S[i] \in ConvexHull(S)$}
   \STATE $PointCircumcenters[i].append(p_{ghost})$
   \ENDIF
   \STATE $PointCircumcenters[i].append(C[S[i]])$.
   
   \ENDFOR

   \IF{multiple of $r$ steps of optimization passed}
   \STATE $T$, $ConvexHull(S)$ = $DT(S)$
   \ENDIF
   \UNTIL{$m$ steps is done}
\end{algorithmic}

\end{algorithm}

\begin{table}
    \centering
    \begin{tabular}{|l|c|} \hline 
         Delaunay triangulation&$O(NlogN)$\\ \hline 
 Calculating circumcenters&$O(N)$\\ \hline 
 Finding border edges&$O(N)$\\ \hline 
 Processing ghost points&$O(N)$\\ \hline 
 Composing $PointCircumcenters$ map&$O(N)$\\ \hline\end{tabular}
    \caption{Algorithmic complexity of the main parts of the method. The complexity of Delaunay triangulation construction is dominating the other parts. During the process of optimization we do retriangulation every $r$ steps. In the worst case we do retriangulation every step, so we have $O(m~NlogN)$ for the method.}
    \label{tab:complexities}
\end{table}
\textbf{Complexity.} Algorithmic complexities of the main parts of the method shown in table \ref{tab:complexities}.
There are $O(N)$ triangles in $DT(S)$.
To find boundary edges we pass though all of them and find those with a single adjacent triangle.
This is $O(N)$ operation because in a Delaunay triangulation the worst case for the number of border edges is then $S = ConvexHull(S)$, which is still linear.
Each ghost point corresponds to a border edge so it's linear too.
Composing $PointCircumcenters$  map is linear because it includes only indexing operation on the pre-computed circumcenters and ghost-points.
After cancelling out the minor terms, the worst-case complexity of the method is $O(m~NlogN)$.

\section{Applications}

\subsection{Differentiable geometry of the Voronoi tessellation}

\begin{figure*}[hbt!]
    \centering
    \includegraphics[width=0.8\linewidth]{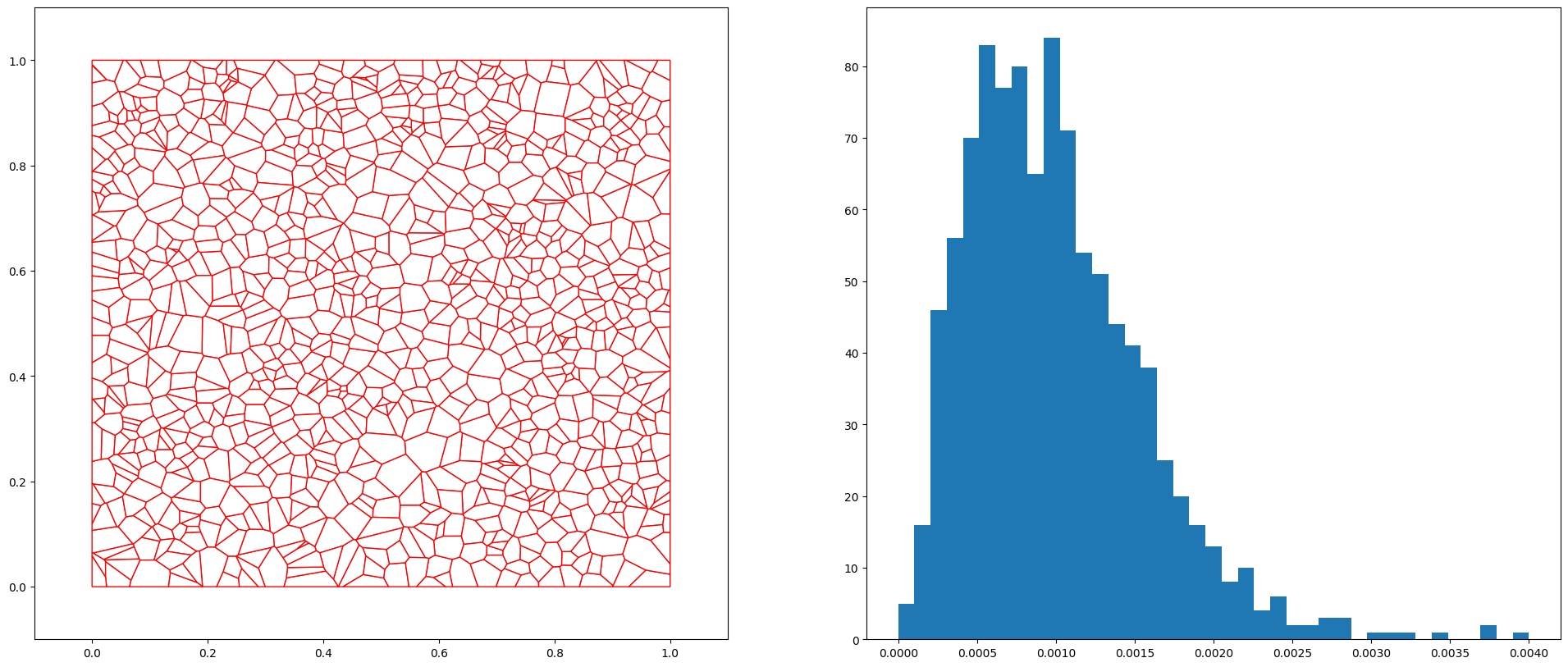}
    \includegraphics[width=0.8\linewidth]{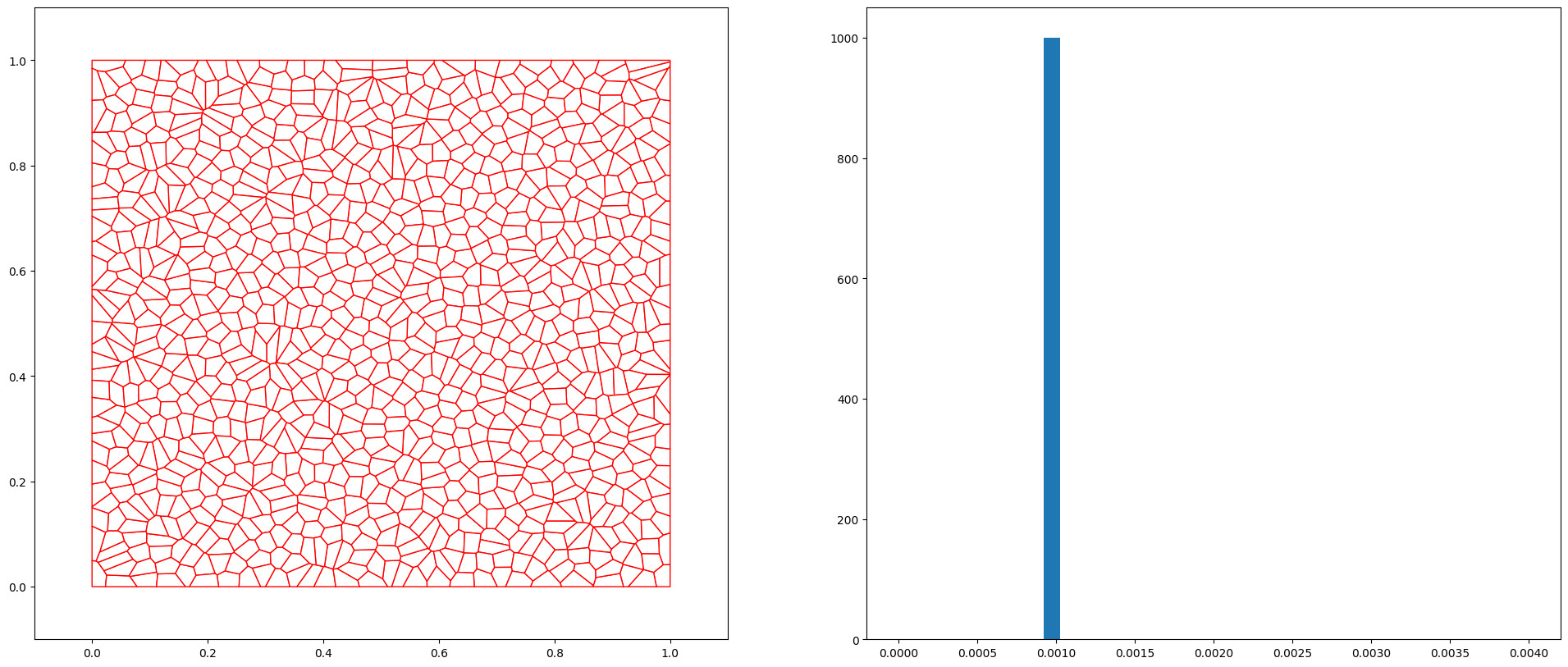}
    \caption{We set the border for the Voronoi tessellation as a unit square. Initially points randomly placed inside the border. The site points are not shown for visual concern. Only the Voronoi edges shown in red. The top-left figure shows the initial Voronoi tessellation. The initial areas of the Voronoi regions differ. The top-right figure shows the distribution of the areas of the Voronoi regions at the start. We optimize the placement of the site points $S$ in order to reduce the variance of the areas of the Voronoi regions. We do 1400 optimization steps using the Adam optimizer with the learning rate 0.001. Bottom-left figure shows the placement of the site points after the optimization. The bottom-right figure shows the distribution of the areas of the Voronoi regions after the optimization. The distribution converges to the theoretical value: unit square with 1000 points means 0.001 square unit per Voronoi region.}
    \label{bounded_optimization}
\end{figure*}

\textbf{Lengths of Voronoi edges}
Voronoi edges connect Voronoi points. 
Each Voronoi edge corresponds to an edge in $DT(S)$.
Non-border edges are adjacent to two triangles.
We simply take their circumcenters and calculate the distance between them.
For border edges we connect the circumcenter of the single adjacent triangle with a ghost point.
We process all edges in $DT(S)$ and calculate the length of the Voronoi edges simply by using edge-triangles, triangles-circumcenters adjacency.

\textbf{Areas of the Voronoi regions.}
To calculate the areas of the Voronoi regions we use the Shoelace formula \cite{braden1986surveyor}:
\begin{align}
A &= \frac 1 2 \sum_{i=1}^n (x_iy_{i+1}-x_{i+1}y_i) &= \\
&=\frac 1 2 \Big(x_1 y_2- x_2 y_1 +x_2 y_3- x_3 y_2+\cdots +x_ny_1-x_1y_n\Big) \label{shoelace}
\end{align}
The Shoelace formula operates by dividing the polygon into triangles and subsequently calculating the area of each triangle using the cross product of two adjacent sides. 
The summation of all triangle areas provides the total area of the polygon. 
The absolute value is taken to ensure that the result is positive, regardless of the order in which the vertices are given.
We implement the formula in a differentiable way and use its output to calculate the loss function which is based on geometry of $V(S)$.

\textbf{Bounded auto-differentiable Voronoi tessellation.}
Most of the practical tasks involving Voronoi diagrams require setting a boundary.
The examples of such tasks are: physical simulation using Finite Element Method, tasks related to Material Science and crystallography.
Here we present the extension of our method for the bounded case.

Let $B$ be an convex $n$-$gon$ representing a boundary.
In the bounded case we construct $V(S)$ the same way but $B$ affects $VR(S)$ and $VE(S)$.
We assume that $ConvexHuss(S) \in B$.
Therefore, we can clip $VR(S)$ and $VE(S)$ after building the regular $V(S)$.
We focus on utilization of differentiable version of the Sutherland–Hodgman algorithm \cite{differentiable_clipping}, well-established method in computational geometry for polygon clipping.
The algorithm works by iterating through each edge of the clipping window and clipping the polygon against that edge.
For each edge, the algorithm determines which vertices of the polygon lie inside or outside of the clipping region.
The intersection points between the polygon edge and the clipping edge are then computed, and these points are added to the output polygon. 
The algorithm continues this process for each edge of the clipping window until all edges have been processed.
The resulting clipped polygon is a subset of the original polygon that lies within the clipping window.
The vertices can occasionally extend beyond the predefined boundary but in our current implementation we don't restrict this behaviour.
Our focus is primarily on the intrinsic parameters such as areas of cells and lengths of edges that can be optimized even the vertex is outside of the bounded area.

\subsection{Minimizing the variance of the Voronoi region's areas}

The primary objective of this experiment is to minimize the areas of Voronoi regions generated from a given set of points. By achieving a lower variance in the areas of $VR(S)$, we aim to obtain a more uniform spatial distribution, which is crucial for various applications, for example, telecommunications.

\begin{figure}[hbt!]
\begin{center}
\includegraphics[width=\columnwidth]{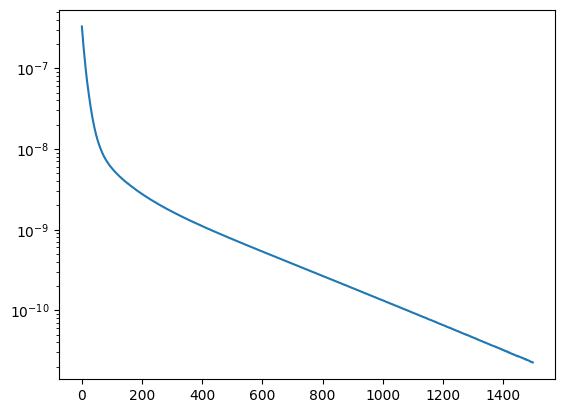}
\caption{We solve the optimization problem described in \ref{optimization}}. We use Adam optimizer with the learning rate 0.001. We do retriangulation on every step, i.e. $r$ = 1. The optimization process shows stable convergence. The optimized variance converges almost to zero.
\label{loss_curve}
\end{center}

\end{figure}

\textbf{Methodology}.
The area of each Voronoi region is calculated using the Shoelace formula (\ref{shoelace}), as described earlier. Let $A_0, ..., A_N$ be the areas of Voronoi regions on the current step of optimization. We use the variance of $A_0, ..., A_N$ as a loss function:
\begin{align}
L(V(S)) = \frac{1}{N} \sum_{i} (A_i - \bar{A})^2,
\end{align}
and solve the following optimization task:
\begin{align}
argmin_S~L(V(S)).
\label{optimization}
\end{align}
We adjust the positions of points $S$ to achieve the minimum variance.
In the unbounded Voronoi case we use mask to filter out infinite Voronoi regions from the loss function on each step of the optimization process.
Figure~\ref{unbounded_optimization} illustrates the optimization results for unbounded case, showcasing the effectiveness of our approach.
We optimized 1000 points in this experiment.
The method shows stable convergence.
The loss curve for the described example is shown in the Fig.~\ref{loss_curve}.

The results of our optimization with a presence of boundary are presented on Figure \ref{bounded_optimization}. As before, we used 1000 points and adjusted their positions during optimization. Theoretical minimum of variance in this case is exactly zero and optimal area of a Voronoi region is 0.001. The ending histogram shows great correspondence of obtained results with theoretical values.

\subsection{Optimizing Hospital Locations}

\textbf{Problem Statement}
Given a set of N hospitals, each with a capacity $C_i$, and a spatial population density function $\rho(x, y)$, the objective is to optimally position these hospitals such that their efficiencies, defined as the ratio of hospital capacity to the integral of population density over the corresponding Voronoi cell, are close to unity. This efficiency parameter, denoted as $W_{\text{eff}}$, is given by

\begin{equation}
W_{\text{eff}, i} = \frac{C_i}{\int_{VR(p_i, S)} \rho(x, y) \, dx dy}
\end{equation}

where $VR(p_i, S)$ represents the Voronoi cell corresponding to the $i$-th hospital. The efficiency parameter $W_{\text{eff}, i}$ varies in the range $[0, \infty)$, where values lower than 1 indicate hospital overload, values greater than 1 indicate underload, and the optimal value is 1.

The problem is to minimize the mean squared error (MSE) loss of the efficiencies of all hospitals with respect to the optimal value of 1. Mathematically, the optimization problem can be formulated as:
\begin{equation}
\text{L} = \quad \frac{1}{N} \sum_{i=1}^{N} \left( W_{\text{eff}, i} - 1 \right)^2
\end{equation}
Let's consider a task where we have continuous density function defined over a region. 
Let the function be $\rho(x, y) = sin(10x) + sin(10y) + 2$. 
Let $\rho$ be defined over the $B=[0, 1]^2$.
We assume that people visit the closest hospital based on the euclidean distance.
The total population is the integral over the unit square:
\begin{equation}
\iint\limits_{B} [sin(10x) + sin(10y) + 2] ~dxdy \approx 2.3678.
\end{equation}

\begin{figure*}[hbt!]
    \centering
    \includegraphics[width=0.8\columnwidth]{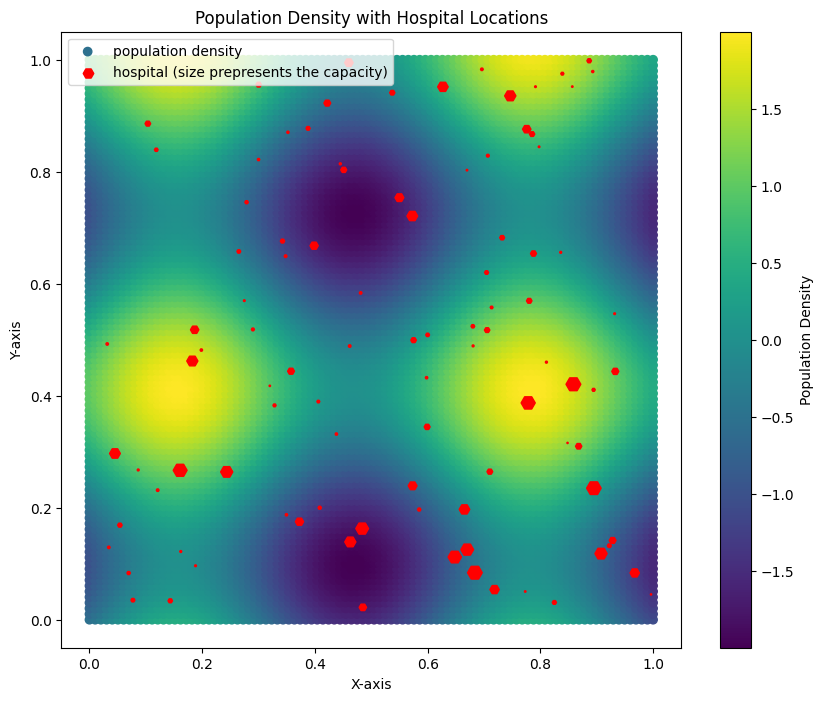}
    \includegraphics[width=0.8\columnwidth]{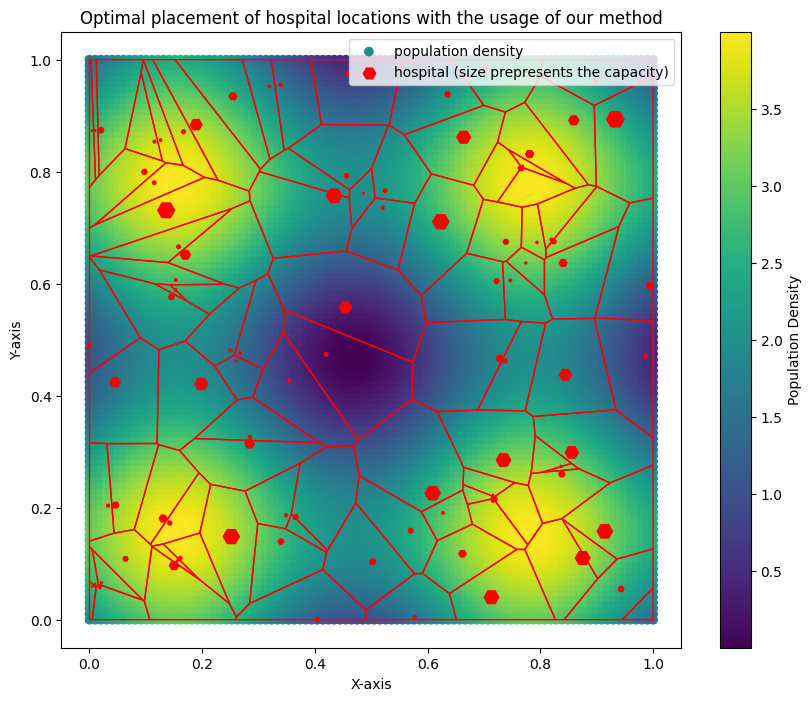}
    \caption{We solve the optimal placement problem of hospitals which have different capacities. We assume that people visit the closest hospital based on the euclidean distance. We define the continuous function over the region and optimize the placement of the hospitals so that the population of each Voronoi cell is exactly the capacity of the hospital it corresponds to.  The size of the red marker corresponds to the capacity of the hospital. We use Adam optimizer with the learning rate 0.001.}
    \label{hospitals}
\end{figure*}

Let define the capacities so that $\sum^N_i C_i = 2.3678$.
The view of the region is presented in Fig.~\ref{hospitals}.
To calculate the integral we use Monte-Carlo estimation of the density function in the Voronoi points of the Voronoi region.

Fig.~\ref{hospitals} (left) shows the results of the optimization. The Voronoi regions correspond to the metric - integral over the Voronoi region equals the capacity of the corresponding site point.

\section{Conclusion}

We presented a new method for auto-differentiating of the Voronoi tessellation.
The key idea is to use AD framework when constructing a Voronoi tessellation.
We use non-differentiable Delaunay triangulation for adjacency information.
Then we use this adjacency information for differentiable construction of the Voronoi tessellation leveraging Voronoi-Delaunay duality.
We make infinite Voronoi regions finite by adding far distant ghost points. 
To find these points we use the properties of the Delaunay triangulation and the convex hull of the site points.
We provided the pseudocode for the method and made the algorithmic complexity assessment.
The geometric parameters of Voronoi tessellation such as lengths of edges and areas of regions can be computed in differentiable way.
It allows using our approach in practical tasks.
In our paper we solved the problem of minimizing variance of the Voronoi regions' areas. 
Our method demonstrated stable convergence and loss almost approached the theoretical minimum.
The developed approach can be particularly useful for solving problems where we need to obtain the gradients of the output with respect to the input through the Voronoi construction.
Such problems include solving inverse problems across a broad range of disciplines.
Overall, our method offers a versatile and efficient approach for handling Voronoi diagrams in various scientific and engineering tasks.

\bibliography{main}
\bibliographystyle{icml2023}

%%%%%%%%%%%%%%%%%%%%%%%%%%%%%%%%%%%%%%%%%%%%%%%%%%%%%%%%%%%%%%%%%%%%%%%%%%%%%%%
%%%%%%%%%%%%%%%%%%%%%%%%%%%%%%%%%%%%%%%%%%%%%%%%%%%%%%%%%%%%%%%%%%%%%%%%%%%%%%%
% APPENDIX
%%%%%%%%%%%%%%%%%%%%%%%%%%%%%%%%%%%%%%%%%%%%%%%%%%%%%%%%%%%%%%%%%%%%%%%%%%%%%%%
%%%%%%%%%%%%%%%%%%%%%%%%%%%%%%%%%%%%%%%%%%%%%%%%%%%%%%%%%%%%%%%%%%%%%%%%%%%%%%%
\newpage
\appendix
\onecolumn
\section{You \emph{can} have an appendix here.}

You can have as much text here as you want. The main body must be at most $8$ pages long.
For the final version, one more page can be added.
If you want, you can use an appendix like this one, even using the one-column format.
%%%%%%%%%%%%%%%%%%%%%%%%%%%%%%%%%%%%%%%%%%%%%%%%%%%%%%%%%%%%%%%%%%%%%%%%%%%%%%%
%%%%%%%%%%%%%%%%%%%%%%%%%%%%%%%%%%%%%%%%%%%%%%%%%%%%%%%%%%%%%%%%%%%%%%%%%%%%%%%

\end{document}